\documentclass[aip,jcp,reprint]{revtex4-1}
\usepackage{graphicx,amsmath}
\begin{document}

\title{Entropic pressure in lattice models for polymers}
\author{Yosi Hammer}
\email{Email: hammeryosi@gmail.com}
\author{Yacov Kantor}
\affiliation{Raymond and Beverly Sackler School of Physics and
Astronomy, Tel Aviv University, Tel Aviv 69978, Israel}
\date{\today}

\begin{abstract}
In lattice models local pressure on a surface is derived from the
change in the free energy of the system due to the exclusion of a certain
boundary site, while the total force on the surface
can be obtained by a similar exclusion of all surface sites. In these
definitions, while the total force on the surface of a lattice system matches the force measured in a continuous system,  the local pressure does not. Moreover, in a lattice system, the sum of the local pressures is not equal to the total force
as is required in a continuous system. The difference is caused by correlation between occupations
of surface sites as well as {\em finite} displacement of surface elements 
used in the definition of the pressures and the force. This problem
is particularly acute in the studies of entropic pressure of polymers
represented by random or self-avoiding walks on a lattice.
We propose a modified expression for the local pressure which
satisfies the proper relation between the pressure and the total
force, and show that for a single ideal polymer in the presence of
scale-invariant boundaries it produces quantitatively correct
values for continuous systems. The required correction to the
pressure is {\em non-local}, i.e., it depends on long range
correlations between contact points of the polymer and the surface.
\end{abstract}

\maketitle

\section{Introduction} \label{sec:intro}
In mechanical systems the pressure is a scalar quantity related to
the diagonal elements of the stress tensor,\cite{MortonE.Gurtin1981}
and in simple homogeneous isotropic equilibrium systems, such as fluids,
it is the main property characterizing the system.\cite{Pathria1996} In
non-homogeneous systems in equilibrium, it is often convenient to consider
 {\em local} stresses inside the system, where they characterize the
momentum transfer, both kinetic momentum and the interaction forces
between particles. (See Refs. \cite{Heinz05,Lion12} and references therein.)
From a kinetic point of view, when classical particles are confined by smooth rigid surfaces (walls),
they undergo specular reflections from the walls, leading to a force
locally perpendicular to the walls. The average of this force per
unit area is the local pressure
\begin{equation} \label{eq:hardWallPress}
P(\vec{x})=k_BTn(\vec{x}),
\end{equation}
where $k_B$ is the Boltzmann constant and $T$ is the temperature,
while  $n(\vec{x})$ is the mean local density of particles in contact
with the wall at position $\vec{x}$. This ideal-gas-like local expression
is unaffected by the interactions between the particles,
and relies solely on the `hard' interaction between each particle
and the wall, where the potential changes from 0 to $\infty$.
In statistical mechanics
$P(\vec{x})$ is the \textit{ensemble averaged} force per unit area acting
on the boundary. In the canonical ensemble such local force is the derivative
of the free energy of the system with respect to a local displacement of
the boundary perpendicular to itself.\cite{Breidenich2007,Bickel2001,Hammer2014}

\begin{figure}
\includegraphics[width=8cm]{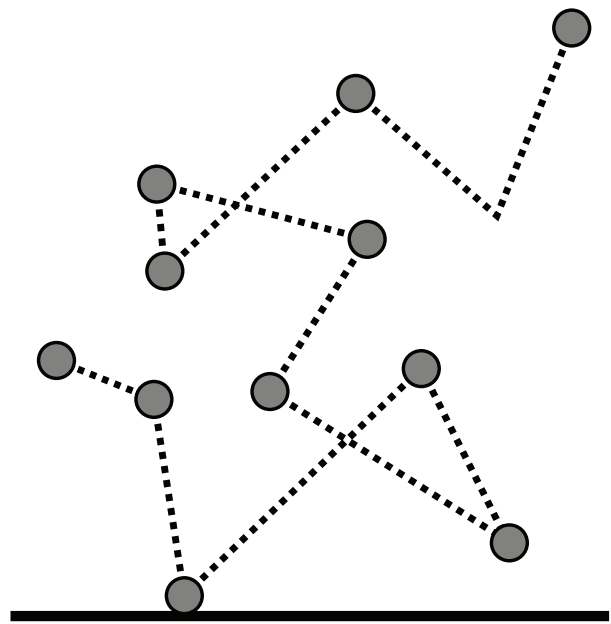}
\caption{A polymer composed of hard spheres (circles) connected
by springs (dashed lines) near the wall of a large box (solid line).
Even though there are several monomers in the polymer, only one is in
contact with the wall.}
\label{fig:BeadsAndSprings}
\end{figure}

Consider a single polymer modeled as a chain of $N+1$ hard spheres
connected by $N$ springs (Fig. \ref{fig:BeadsAndSprings}), placed in a box with flat hard walls. 
The probability of finding a monomer within a small distance $\epsilon$ from the boundary is linear in $\epsilon$, while the probability of finding {\em two} monomers is proportional to $\epsilon^2$. Thus, when the wall is shifted by an infinitesimal distance, only one monomer makes contact with the wall and the polymer touches the wall at a single point. The local nature of the pressure is reflected in Eq.\eqref{eq:hardWallPress}, where the pressure depends on the monomer density at a specific point $\vec{x}$. Note that this density is affected by the interaction between the monomers in the polymer.

The `beads and springs' model is a {\em discrete} model defined in a {\em continuous} space. If we take the $N\rightarrow\infty$ limit and at the same time take all the microscopic length scales to zero (the radius of the spheres and the average length of the springs), so that the average end-to-end distance $R$ of the polymer remains finite, we arrive at a \textit{continuous} polymer model.\cite{Doi1986} When we place a continuous polymer near a repulsive wall, the density of monomers on the wall vanishes. The entropic pressure of the polymer on the wall at the point $\vec{x}$ is determined by the rate of change of the monomer density close to $\vec{x}$ in the direction perpendicular to the wall.\cite{Hammer2014,Bickel2001,Eisenriegler1993} In the above models, which are defined in continuous space, the total force is found by integrating the pressure over the surface, i.e.,
\begin{equation} \label{eq:hardWallTotalForce}
\vec{F}=\int Pd\vec{S},
\end{equation}
where $d\vec{S}$ is a vector whose size is that of a surface element, and it is perpendicular to the wall.

\begin{figure}
\includegraphics[width=8cm]{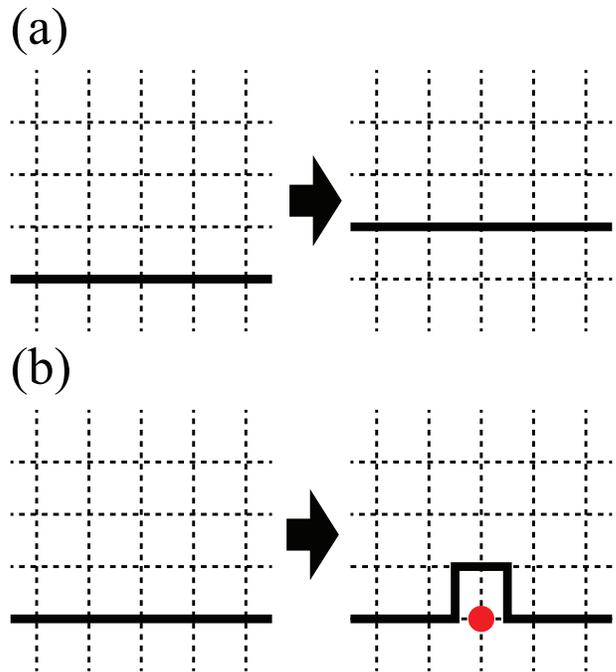}
\caption{(Color online) Discrete changes on a square lattice. (a) The wall is shifted by one lattice constant. (b) The lattice site marked by the red circle is excluded.}
\label{fig:LatticeChanges}
\end{figure}

A random walk (RW) on a lattice is often used as a model for an ideal polymer, and a self-avoiding walk (SAW) is used as a model for a polymer in good solvent.\cite{Gennes1979} Using lattice models in polymer simulations allows qualitative treatment of universal features of larger systems at the expense of quantitative agreement with real systems.\cite{Telyanskii2004} In this work we propose a method to recover some of this quantitative agreement.
In lattice systems the statistical mechanics approach to the local pressure is reduced to the calculation of discrete changes of the system.\cite{Dickman1987,Jensen2013,VanRensburg2013,Gassoumov2013} Consider a RW or a SAW on a lattice confined to a large box. The walk is allowed to visit the sites on the walls of the box but not to cross them. The force $F$ acting on one flat wall is found by calculating the change in the free energy of the system $\Delta\mathcal{F}$ that results from excluding all the sites along the wall, thus moving the wall by a finite distance $\Delta h$, i.e.,
\begin{equation} \label{eq:latticeForce}
F\Delta h=\Delta \mathcal{F}.
\end{equation}
If the wall is a plane (in space dimension $d=3$) or a line (in $d=2$) and it is shifted along one of the axes of a hypercubic lattice, as in Fig. \ref{fig:LatticeChanges}a, $\Delta h$ is simply the lattice constant $a$. Similarly, it is natural to derive the pressure from the change $\Delta\mathcal{F}(\vec{x})$ in the free energy of the system resulting from the exclusion of a lattice site $\vec{x}$,
 \begin{equation} \label{eq:latticePress}
 P(\vec{x}) \Delta h \Delta S =\Delta\mathcal{F}(\vec{x}),
 \end{equation}
where $\Delta S$ is the surface element associated with the site $\vec{x}$. In the simple example depicted in Fig. \ref{fig:LatticeChanges}b, $\Delta S$ is the lattice constant, and $\Delta S\Delta h = a^2$.

\begin{figure}
\includegraphics[width=8cm]{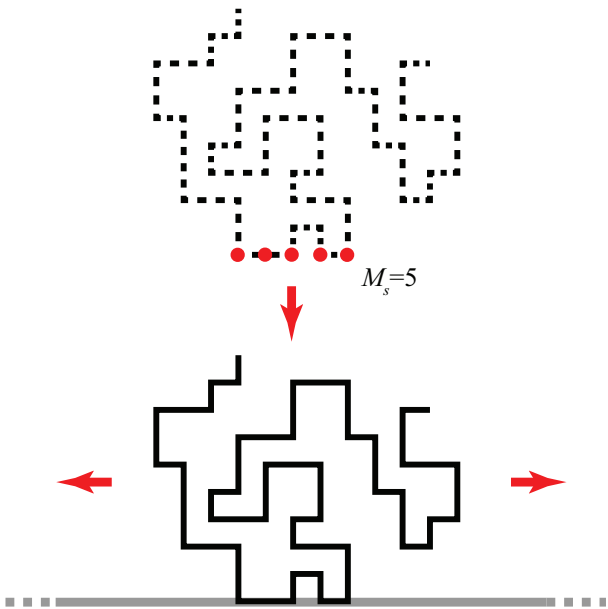}
\caption{A SAW on a square lattice brought into contact with the wall of a large box. The system is symmetric to translations along the wall. The walk touches the wall at $M_s=5$ different sites.}
\label{fig:WalkNearWallHomo}
\end{figure}

Unlike in polymer models in continuous space, in the lattice polymer models there can be more than one monomer in contact with the wall (Fig. \ref{fig:WalkNearWallHomo}). More than one site along the wall can be occupied, and in the case of a RW more than one monomer can occupy a single site. For this reason, Eq.~\eqref{eq:hardWallTotalForce} does not hold in the lattice model, i.e., \cite{Jensen2013}
 \begin{equation} \label{eq:latticeForcePressIneq}
 \vec{F}\neq\int Pd\vec{S}.
 \end{equation}
Thus the local pressure defined in this way is not a good representation of what we mean by pressure in continuous systems. We seek to correct this situation, and define a local pressure in a discrete system that is numerically as close as possible to the pressure measured in continuous systems and results in the correct total force after integration.

\section{General formalism for lattice polymer problems}
Consider a polymer modelled as a RW or SAW of $N$ steps on a hypercubic lattice confined to a certain region of space. We study the pressure and the total force on one of the walls of the allowed region. The configurational part of the free energy of the system is
\begin{equation} \label{eq:latticeFreeEnergy}
\mathcal{F} = -k_BT\ln\mathcal{N}_t,
\end{equation}
where $\mathcal{N}_t$ is the total number of allowed configurations. We denote by $\mathcal{N}_c$ the number of configurations in which the polymer touches the wall at least at one site. When the wall is moved by one lattice constant the total number of available configurations is reduced by $\mathcal{N}_c$, so that the change in the free energy of the system is
\begin{equation} \label{eq:latticeFreeEnergyChange}
\Delta\mathcal{F}=-k_BT\ln\left(1-\frac{\mathcal{N}_c}{\mathcal{N}_t}\right).
\end{equation}
It is convenient to define a dimensionless force variable $\hat{F}\equiv\mathcal{N}_c/\mathcal{N}_t$, so that the total force on the wall is
\begin{equation} \label{eq:FandFhat}
F=-\frac{k_BT}{\Delta h}\ln\left(1-\hat{F}\right).
\end{equation}
Frequently, $\hat{F}\ll1$ and we can expand Eq.~\eqref{eq:FandFhat} and see that $\hat{F}$ is indeed proportional to the total force, i.e., $F\approx(k_BT/\Delta h)\hat{F}$.
We repeat this process with respect to the change in the free energy of the system $\Delta\mathcal{F}(\vec{x})$ associated with the exclusion of the lattice site $\vec{x}$ from the available  sites for the walk. If we denote by $\mathcal{N}_{\vec{x}}$ the number of walks that touched the surface at the point $\vec{x}$ at least once, we arrive at
\begin{equation} \label{eq:PandPhat}
P(\vec{x})=-\frac{k_BT}{\Delta h \Delta S}\ln\left(1-\hat{P}(\vec{x})\right),
\end{equation}
where the dimensionless pressure $\hat{P}(\vec{x})\equiv\mathcal{N}_{\vec{x}}/\mathcal{N}_t$.  \cite{Bickel2001,Gassoumov2013,VanRensburg2013,Brum14} Practically always $\hat{P}(\vec{x})\ll1$, and we can expand Eq.~\eqref{eq:PandPhat} to get $P(\vec{x})\approx(k_BT/\Delta h \Delta S)\hat{P}(\vec{x})$.

\section{Force and pressure of a lattice polymer in a large box} \label{sec:homo}
When the polymer is confined to a large box (the size of the box is much larger than the average size of the polymer) the system can be regarded as homogeneous, and to a good approximation, invariant with respect to translations along the wall. In this case the relation between the pressure and the total force can be derived from the analysis of lattice polymers in free space (no boundaries). Any configuration in free space can be created by taking a walk $s$ starting from the origin and moving it as a rigid unit to a new position $\vec{x}_0$ (Fig. \ref{fig:WalkNearWallHomo}). Thus each walk $w$ is identified by its shape $s_w$ and its starting position $\vec{x}_{0,w}$. Suppose we wish to study the pressure and force between walks in a large box, and one of the walls which is defined by $x_i=0$, where $x_i$ is the $i$ spatial coordinate in the system. For each shape $s$ we can identify $x_{i,\text{min}}$, the minimal value of $x_i$ along the walk, and define $M_s$ to be the number of different minimal sites of the walk in the direction of $i$ (Fig. \ref{fig:WalkNearWallHomo}). When a walk $w$ is in contact with the wall, i.e., $x_{i,\text{min}}=0$, it touches the wall at precisely $M_{s_w}$ sites. Note that this is independent of the starting position of the walk along the wall, and only depends on it's shape.  Due to the symmetry of the system with respect to translations parallel to the large wall, the number of configurations in which the polymer is in contact with the wall is 
\begin{equation} \label{eq:Nc}
\mathcal{N}_c=\mathcal{N}_s \cdot \mathcal{N}_b,
\end{equation}
where $\mathcal{N}_s$ denotes the number of shapes the polymer can take (for a RW on a $d$-dimensional hypercubic lattice $\mathcal{N}_s=(2d)^N$) and $\mathcal{N}_b$ is the number of sites on the wall.
 The number of configurations that touched the wall at a certain site $\vec{x}$ at least once is independent of $\vec{x}$ due to the same translational symmetry, and can be expressed as
\begin{equation} \label{eq:Nx}
\mathcal{N}_{\vec{x}}=\sum_sM_s.
\end{equation}
From  Eqs.~\eqref{eq:Nc}, \eqref{eq:Nx} and the definition of $\hat{P}$ and $\hat{F}$ we arrive at
\begin{eqnarray}
\sum_{\vec{x}}\hat{P}(\vec{x})&=&\frac{\mathcal{N}_b}{\mathcal{N}_t}\sum_sM_s=\hat{F}\frac{1}{\mathcal{N}_s}\sum_sM_s \notag\\
&=&\langle M\rangle \hat{F}.
\end{eqnarray}
We see that when the polymer is moving freely in a large box, the ratio between
the total force and the integral of the pressure on the surface of the wall is
equal to the average $\langle M\rangle$ taken over possible shapes of the walk.
It is interesting to note that this average can be measured without the presence
of the confining box, and it is in fact a property of polymers in free space.
The technique of using
walks in free space to study properties of walks in the presence of
boundaries is sometimes referred to as CABS (confinement analysis from bulk
structure), and has been discussed and used in several recent
works. \cite{Wang2008,Wang2008a} By numerically examining a RWs of
$N=10^5$ steps on a square lattice we find $\langle M\rangle\approx2.78$.
Thus, for RWs on a square lattice, when the walk moves freely in a large
confining box with a flat wall of surface $S$, then
the pressure on the wall and the total force $F$ acting on it are related
by $PS=2.78F$. In fact, as will be discussed later, even in more complicated, non-homogeneous systems, it is often sufficient to divide the pressure by the constant $\langle M\rangle$ computed in free space.

\section{Force and pressure of a lattice polymer in a non-homogeneous case} \label{sec:nonHomo}
We now examine the relation between the local pressure and the total
force applied on the wall for a non-homogeneous system. Consider, for example,
a lattice polymer anchored at a site $\vec{x}_0$ near an infinite wall (Fig. \ref{fig:geometries}a). The polymer is confined to the half space defined by the wall. As before, it can visit the sites on the boundary but is not allowed to cross it.
Let us denote by $w$ a certain configuration of the polymer that {\em visited} the boundary wall. Also, let us define a variable $m_{w}(\vec{x})$ that is equal to one if the walk $w$ visited the site $\vec{x}$ on the wall, and zero otherwise (so that $\sum_{\vec{x}}m_{w}(\vec{x})=M_{w}$, where $M_w=M_{s_w}$ defined above). The total number of walks that visited the wall can be written as a sum over the configurations $w$ that touched the wall, i.e,
\begin{equation} \label{eq:NcNonHomo}
\mathcal{N}_c=\sum_{w}1=\sum_{w}\sum_{\vec{x}}\frac{m_{w}(\vec{x})}{M_{w}}.
\end{equation}
Combining Eq.~\eqref{eq:NcNonHomo} with the definitions of $\hat{F}$ and $\hat{P}$, we find that
\begin{eqnarray} \label{eq:FhatAndPhatNonHomo}
\hat{F}&=&\frac{1}{\mathcal{N}_t}\sum_{\vec{x}}
\sum_{w}\frac{m_{w}(\vec{x})}{M_{w}} \\
&=&\sum_{\vec{x}}\frac{\mathcal{N}_{\vec{x}}}{\mathcal{N}_t}
\frac{1}{\mathcal{N}_{\vec{x}}}\sum_{w_{\vec{x}}}
\frac{m_{w_{\vec{x}}}(\vec{x})}{M_{w_{\vec{x}}}} \notag\\
&=&\sum_{\vec{x}}\hat{P}(\vec{x})\left\langle M^{-1}\right\rangle_{\vec{x}},\notag
\end{eqnarray}
where $w_{\vec{x}}$ is a configuration that touched the point $\vec{x}$ on
the wall. We denote by $\langle\rangle_{\vec{x}}$ a \textit{conditional}
average with respect to only the walks that visited the site $\vec{x}$.
Note that the switch from summation over $w$ to the summation over
$w_{\vec{x}}$ in Eq.~\eqref{eq:FhatAndPhatNonHomo} is justified since
for any walk $w$ that did not visit the site $\vec{x}$  we will have
$m_{w}(\vec{x})=0$ and it will not contribute to the sum.
We see from Eq.~\eqref{eq:FhatAndPhatNonHomo} that it is possible to
define a \textit{modified} dimensionless pressure
\begin{equation}
\bar{P}(\vec{x})=\hat{P}(\vec{x})\langle M^{-1}\rangle_{\vec{x}},
\end{equation}
that, when integrated on the boundary wall, will result in the total force acting on the wall. 

The modified definition of the entropic pressure was derived for a {\em single} polymer near a boundary wall. It is straight forward to generalize $\bar{P}$ to a system with multiple polymers. In the latter case $w$ would represent a specific configuration of {\em all} molecules in the system in which at least one monomer contacts the surface and $M_w$ would denote the number of different sites on the surface occupied in $w$. The remainder of the derivation would not be affected.

In the next section, we study the properties of this correction and check whether the modified pressure recovers the local pressure defined in the continuous models.

\section{Polymers on a square lattice anchored near a confining line or sector} \label{sec:squareLattice}

\begin{figure}
\includegraphics[width=8cm]{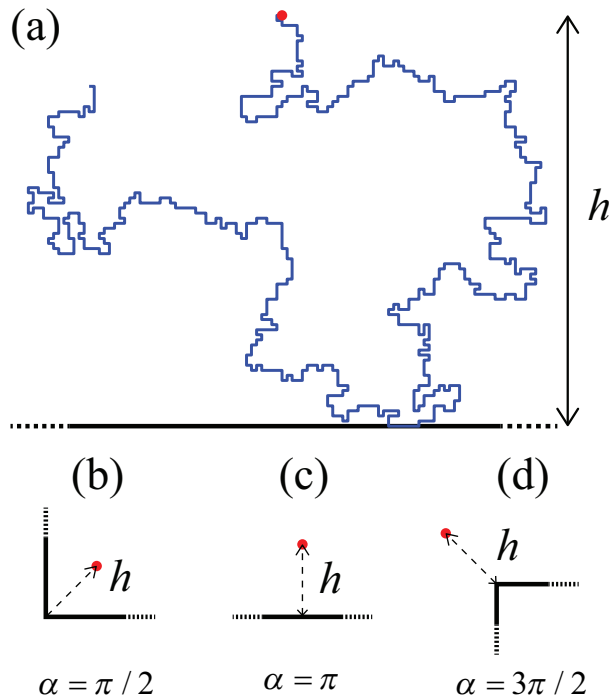}
\caption{(a) (Color online) A SAW on a square lattice in contact with a boundary line. The starting position of the walk (red circle) is at a distance $h$ from the line. (b)-(d) The geometries considered in the force and pressure measurements of section \ref{sec:squareLattice} are sectors with different opening angles $\alpha$.}
\label{fig:geometries}
\end{figure}

We studied the pressure and the force between a RW or  a SAW on a square lattice starting from a site $\vec{x}_0$ near a confining boundary. Three cases are considered for the geometry of the boundary: an infinite line (Fig. \ref{fig:geometries}a,c), a sector of opening angle $\alpha=\pi/2$ (Fig. \ref{fig:geometries}b) and a sector of opening angle $\alpha=3\pi/2$ (Fig. \ref{fig:geometries}d). The starting position of the polymer was taken to be on the symmetry axis of the sector (irrelevant for the infinite line) at a distance $h$ from the boundary.

\subsection{Entropic force measurement}
\begin{figure}[t]
\includegraphics[width=8cm]{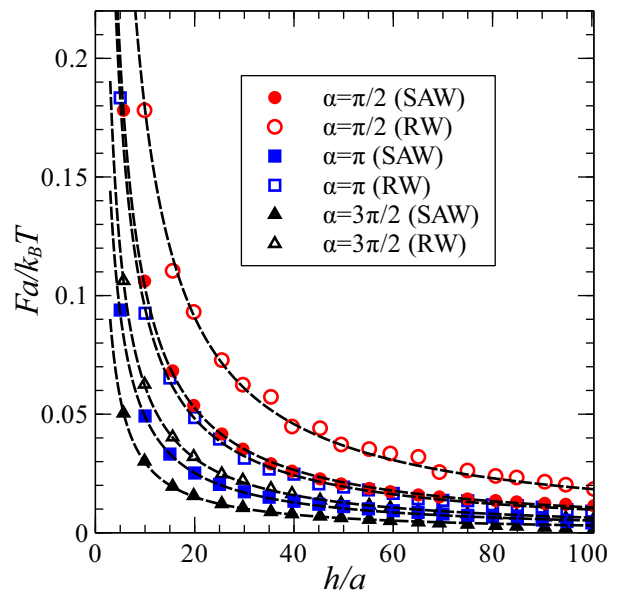}
\caption{(Color online) Entropic force between a RW of $10^5$ steps or a SAW of 512 steps on a square lattice and a scale invariant boundary, shown as a function of the distance $h$ between the starting position of the walk and the boundary in lattice units $a$. The shape of the boundary is a sector with different opening angles $\alpha$ (see legend). The data were fitted to the form in Eq.~\eqref{eq:ForceFit} (dashed lines).}
\label{fig:Force}
\end{figure}

The entropic force between long polymers and scale invariant surfaces was studied in previous papers. \cite{Maghrebi2011,Maghrebi2012} It was shown that for continuous polymers in the limit where the typical linear size of the polymer $R\rightarrow\infty$ the entropic force between the polymer and the surface is of the form
\begin{equation}
F=\mathcal{A}k_BT/h,
\end{equation}
where the force amplitude $\mathcal{A}$ is a universal constant, i.e., it depends on a small number of parameters such as the dimension of the system, the opening angle of the sector and the presence of steric repulsion between monomers. Specifically,
\begin{equation} \label{eq:ForceAmpExact}
\mathcal{A}=\eta_b-\eta_f,
\end{equation}
where $\eta$ is the critical exponent that characterizes the anomalous decay of correlations between monomers.\cite{Cardy1996} We denote by $\eta_f$ the value of this exponent for a polymer in free space (no boundary) and by $\eta_b$ the value of $\eta$ for a polymer anchored to the boundary, i.e., anchored to the confining line or to the tip of the confining sector. It was also shown in refs. \cite{Maghrebi2011,Maghrebi2012} that for RWs, $\eta_f=0$, whereas $\eta_b=\pi/\alpha$, leading to
\begin{equation} \label{eq:RWForceAmp}
\mathcal{A}_{\text{RW}}=\pi/\alpha.
\end{equation}
Cardy and Redner \cite{Cardy1984} found the critical exponents for long SAWs confined to sectors in two dimensions using conformal mapping. From their results,
\begin{equation} \label{eq:SAWFroceAmp}
\mathcal{A}_{\text{SAW}}=\frac{30-5\alpha/\pi}{48\alpha/\pi}.
\end{equation}
In order to demonstrate a measurement of the entropic force, we generated a large number ($\sim10^7$) of RWs of $10^5$ steps on a square lattice, each starting at a distance $h$ from a boundary line or sector as described above. Walks that crossed the boundary were discarded, and among the walks that remained within the allowed space, we counted how many visited the sites on the boundary, thus measuring the ratio $\hat{F}=\mathcal{N}_c/\mathcal{N}_t$. The entropic force is then given by Eq.~\eqref{eq:FandFhat}. The measurement was performed in a similar way for SAWs, where walks of 512 steps were generated using dimerization.\cite{JansevanRensburg2009} The results are presented in Fig. \ref{fig:Force}. 
The form in Eq.~\eqref{eq:ForceAmpExact} for the force is valid only when the distance $h$ from the wall is much greater than any microscopic length scale such as the lattice constant. This is clearly not the case for some of the data presented in Fig. \ref{fig:Force}. 
It has been shown in several works, that to first order, the affect of the microscopic length scale can be taken into account by adding a constant shift $\delta$ to the distance $h$.\cite{Milchev1998,Hsu2004,Teraoka2002,Dimitrov2008} For this reason, we fitted the force measurement data to a function of the form
\begin{equation} \label{eq:ForceFit}
\frac{F(h)}{k_BT}=\frac{\mathcal{A}}{h+\delta},
\end{equation}
where the force amplitude $\mathcal{A}$ and the shift $\delta$ are the fitting parameters. In Table \ref{tab:ForceAmplitudes} we see that the force amplitudes extracted from the measurements are in good agreement with the exact values (Eqs.~\eqref{eq:RWForceAmp} and \eqref{eq:SAWFroceAmp}) for $\alpha=3\pi/2$ and $\alpha=\pi$. There are noticeable discrepancies between the theoretical and the measured values when the polymer is confined inside a sector with $\alpha=\pi/2$, where the polymer is closest to the walls and we can expect that the small length scale will be most significant. Another reason for these discrepancies is the fact that the theoretical force amplitudes were computed for infinitely long polymers, while our walks are finite. The finiteness of the walks leads to a reduction of the force. (see discussion for a finite walk near a line in the next section).  

\begin{table}
\begin{tabular}{c | c | c | c | c}
& \multicolumn{2}{c |}{RW} & \multicolumn{2}{c}{SAW} \\ \hline
$\alpha$ & theory & numerics & theory &  numerics \\ \hline
$\pi/2$ & 2 & 1.85 $\pm$ 0.06 & 1.15 & 1.07 $\pm$ 0.02\\
$\pi$ & 1 & 0.97 $\pm$ 0.02 & 0.52 & 0.52 $\pm$ 0.02\\
$3\pi/2$ & 2/3 & 0.64 $\pm$ 0.02 & 0.31 & 0.31 $\pm$ 0.02\\
\end{tabular}
\caption{Force amplitudes for RWs of $10^5$ steps and SAWs of $512$ steps
on a square lattice anchored near an infinite boundary in the shape of a
sector with opening angle $\alpha$. Theoretical values for the
infinitely long continuous walks are compared with the numerical
values  extracted from Fig. \ref{fig:Force}. The errors represent $95\%$
confidence limits. We did not attempt to estimate systematic errors.}
\label{tab:ForceAmplitudes}
\end{table}

\subsection{Entropic pressure measurement}
\begin{figure}[h!]
\includegraphics[width=8cm]{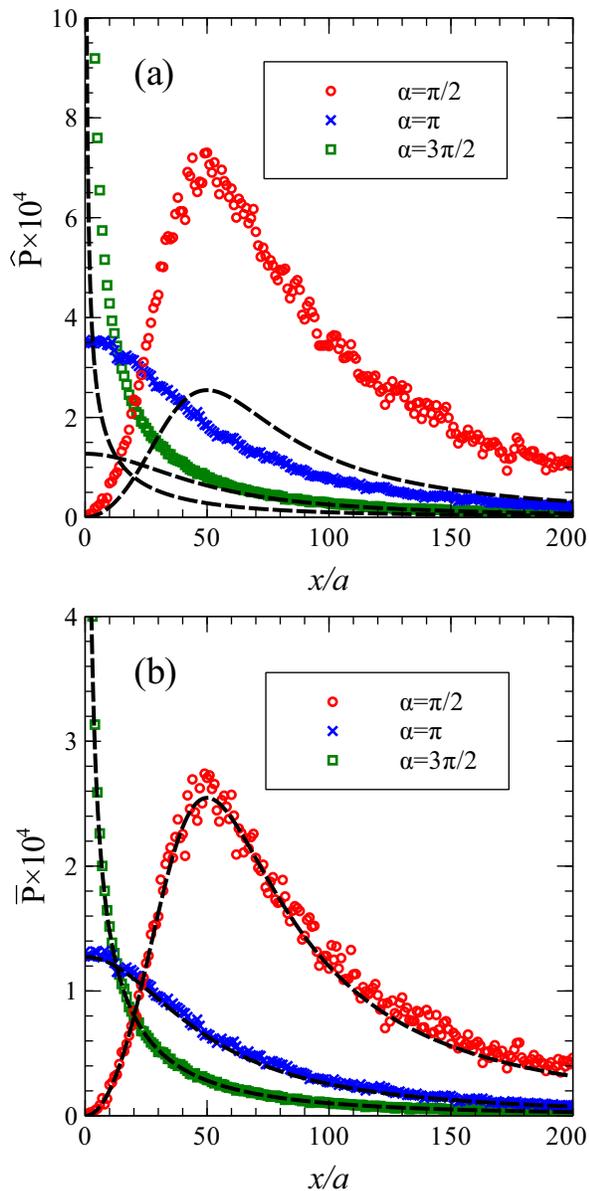}
\caption{(Color online) Entropic pressure of a RW of $10^5$ steps on a square lattice anchored approximately at a distance of 50 lattice constants ($h=50a$) from a scale invariant boundary. (a) The dimensionless pressure $\hat{P}$. (b) The modified dimensionless pressure $\bar{P}$. The dashed lines denote the exact pressure of an infinite continuous ideal polymer (Eq.~\eqref{eq:exactPressure}).}
\label{fig:RWsectorPressure}
\end{figure}
Using the same set of configurations generated for the force calculation, we measured the dimensionless pressure $\hat{P}(x)=\mathcal{N}_x/\mathcal{N}_t$, and the corrected pressure $\bar{P}(x)=\hat{P}(x)\langle M^{-1}\rangle_x$, where $x$ is the distance from the tip of the sector (for the infinite line the starting position of the polymer is directly above the point where $x=0$). We compare these measurements with the exact results obtained in a previous work,\cite{Hammer2014} where it was shown that for a continuous infinitely long ideal polymer, held at one end near a scale invariant repulsive sector, the pressure on the sector is
\begin{equation} \label{eq:exactPressure}
P_e(x)= \frac{\pi}{\alpha^2}\frac{k_BT}{x^2}\frac{1}{1+(h/x)^{2\pi/\alpha}}.
\end{equation}
In Fig. \ref{fig:RWsectorPressure} we present the dimensionless pressure $\hat{P}$ (Fig. \ref{fig:RWsectorPressure}a) and the corrected pressure $\bar{P}$ (Fig. \ref{fig:RWsectorPressure}b) for a random walk confined to a sector. It can be seen that $\bar{P}$ is in much better agreement with $P_e$ (Eq.~\eqref{eq:exactPressure}) than $\hat{P}$. For the majority of values of $x/a$ depicted in Fig. \ref{fig:RWsectorPressure}, the modification factor $\langle M^{-1}\rangle_x$ is almost constant (see the discussion in the next subsection). We could, therefore, obtain very similar results by dividing the reduced pressure by the constant $\langle M\rangle$ found in section \ref{sec:homo}. For these simple surfaces, it is not clear from the data that the more complicated procedure we suggest here of calculating $\langle M^{-1}\rangle_x$ at each point individually is preferable to simple division of the pressure by a constant. However, the modified pressure is guaranteed to result in the correct total force upon integration on {\em any} surface, including more complicated ones that are not homogeneous.

\begin{figure}[t]
\includegraphics[width=8cm]{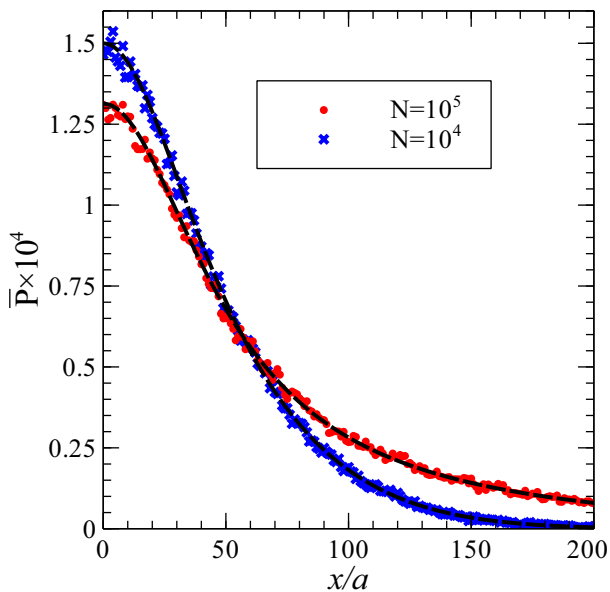}
\caption{(Color online) Entropic pressure of a finite RW on a square
lattice of $N$ steps, anchored at a distance of 50 lattice constants
($h=50a$) from a boundary line. The dashed lines denote the exact
pressure given by Eq.~\eqref{eq:exactPressureFinite}.}
\label{fig:RWFinite}
\end{figure}

A close inspection of the graph reveals that there is still a systematic discrepancy between $\bar{P}$ and $P_e$. The measured pressure $\bar{P}$ is slightly larger than the theoretical pressure $P_e$. The reason is that our walks are finite. In order to demonstrate this, we used the methods described in ref. \cite{Hammer2014} to calculate the pressure of a \textit{finite} continuous ideal polymer in two dimensions anchored at a distance $h$ near an infinite boundary line (a sector where $\alpha=\pi$),
\begin{equation} \label{eq:exactPressureFinite}
P_e(x,R)=2k_BT\frac{h}{R^3}\frac{G_{1,2}^{2,0}\left( \left. \frac{h^2+x^2}{R^2}\right|\begin{array}{c}
-\frac{1}{2}\\-1,0\end{array}\right)}{\pi \, \text{erf}(h/R)},
\end{equation}
where $R=Na^2$, $G_{1,2}^{2,0}$ is the Meijer function
\cite{RyzhikI.S.Gradshteyn2007} and erf is the error function. In
Fig. \ref{fig:RWFinite} we show $\bar{P}$ measured for RWs of
$N=10^4$ and $10^5$ steps. The results are in excellent agreement
with Eq.~\ref{eq:exactPressureFinite}. The dependence of the
pressure on the size of the polymer is also seen in the figure.
For smaller polymers, the pressure at small $x$ is larger in
comparison with the infinite polymer limit, but when $x$ approaches
the size of the polymer the pressure is cut off exponentially. The
exact total force of a finite continuous polymer on the line can be
found by integrating Eq.~\ref{eq:exactPressureFinite}. We find that
the force applied by a finite walk, $F_{\text{finite}}$, approaches
the infinite limit $F_{\text{infinite}}$ from below. For $R\gg h$,
the leading correction $F_{\text{finite}}-F_{\text{infinite}}\sim1/R$.

\begin{figure}
\includegraphics[width=8cm]{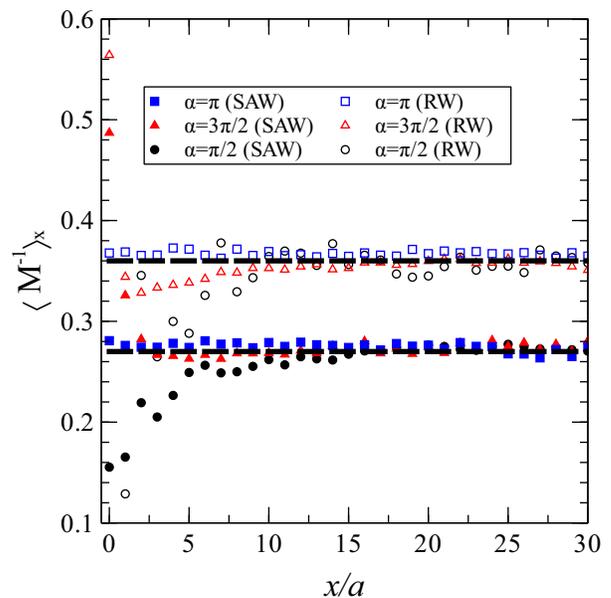}
\caption{(Color online) The correction factor $\langle M^{-1}\rangle_x$
measured for RWs of $10^5$ steps and SAWs of $512$ steps on a square
lattice anchored near a confining sector. The RWs were anchored
approximately at a distance $h=50a$ from the tip of the sector and
the SAWs were anchored at a distance of approximately $h=30a$ from
the tip of the sector (see Fig. \ref{fig:geometries}). For the case
of a boundary line (sector with opening angle $\pi$) the correction
factor is homogeneous (squares). For a sector with $\alpha=\pi/2$
(circles) the correction decays to zero when the corner is approached
(as do the pressure in this case). For a sector with opening angle
$\alpha=3\pi/2$ (triangles), the correction factor increases near the
tip (as does the pressure).}
\label{fig:correctionFactor}
\end{figure}

\begin{figure}[t]
\includegraphics[width=8cm]{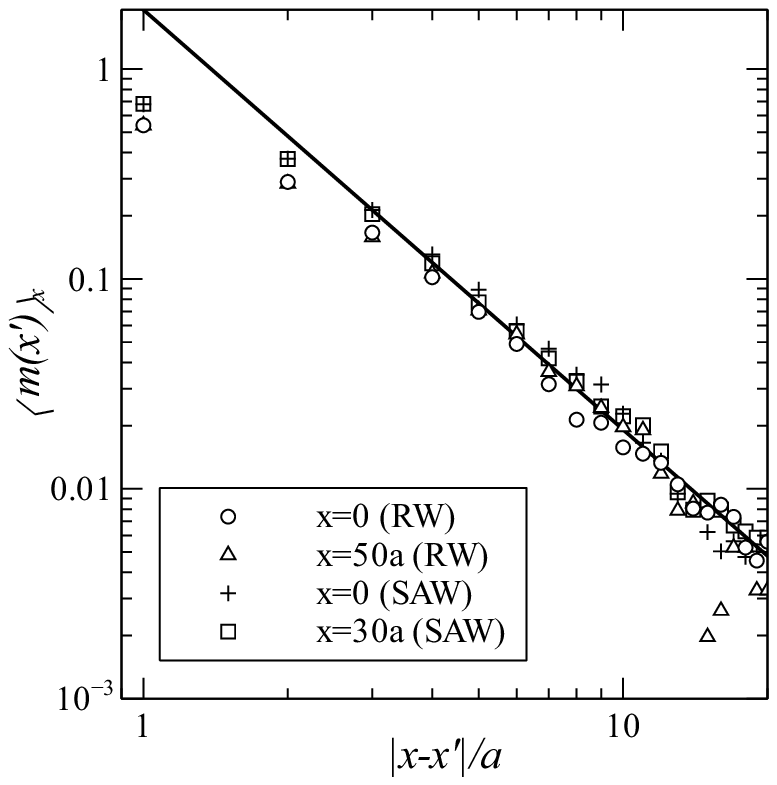}
\caption{The conditional function $\langle m(x')\rangle_x$ versus the
distance $|x-x'|$ between the sites, taken for RWs of $N=10^5$ steps
and SAWs of $N=512$ steps on a square lattice. The starting position of
the RWs was 50 lattice constants above a repulsive boundary line and for
the SAWs it was 30 lattice constants above the line. Apart from the first
two points, the density decays as $|x-x'|^{-2}$ (solid line). The conditional
density is the same for $x=0$ (circles and plus signs) directly beneath
the starting position and for $x=50a$ or $x=30a$ (triangles and squares),
even though the systems are not homogeneous.}
\label{fig:mDecay}
\end{figure}

\subsection{Properties of the factor $\langle M^{-1}\rangle_{\vec{x}}$}
We would like to understand whether the reduction factor $\langle M^{-1}\rangle_{\vec{x}}$ applied to the pressure is a \textit{local} parameter, i.e., it results from the statistics of the system in the vicinity of $\vec{x}$, or rather it depends on long range correlations of the polymer. To achieve this, we note that it can be written in the following way,
\begin{equation} \label{eq:correctionFactorForm2}
\langle M^{-1}\rangle_{\vec{x}}=\left\langle\frac{1}{1+\sum_{\vec{x}'\neq\vec{x}}m(\vec{x}')}\right\rangle_{\vec{x}}.
\end{equation}
From Eq.~\eqref{eq:correctionFactorForm2} we see that the other sites on the boundary affect the correction to the pressure at the site $\vec{x}$ through the conditional function $\langle m(\vec{x}')\rangle_{\vec{x}}$. If this function decays rapidly as the distance $|\vec{x}-\vec{x}'|$ increases (say exponentially), then we could say that the modification of the pressure is local, and depends only on contacts made by the walk in the vicinity of the site $\vec{x}$. In the case of RWs and SAWs on a square lattice confined to a sector, we find that $\langle m(x')\rangle_x$ is not a local function, but decays as a power of $|x'-x|^{-2}$ (Fig. \ref{fig:mDecay}). In fact, it is possible to show that for a RW in $d$ dimensions that touches a ($d-1$)-dimensional boundary plane at a point $\vec{x}$, the function $\langle m(\vec{x}')\rangle_{\vec{x}}$ decays as $|\vec{x}-\vec{x}'|^{-d}$. For SAWs on a square lattice, there seems to be a similar behaviour.
This observation leads us to say that in the 2-dimensional case, the correction we found to the entropic pressure is \textit{non-local}. This property of the entropic pressure in a lattice system makes it fundamentally different from the pressure in continuous systems, discussed in section \ref{sec:intro}. In lattice systems, the idea of a pure local pressure which results in the total force on the surface after integration is no longer valid, since the correction $\langle M^{-1}\rangle_{\vec{x}}$ contains important long range contributions.

Another interesting question is whether the reduction factor applied to the pressure is constant or does it vary with position along the boundary (like the pressure). In Fig. \ref{fig:correctionFactor} we plot $\langle M^{-1}\rangle_{\vec{x}}$ for a RW anchored at a distance $h=50a$ from the sector, and a SAW anchored at a distance $h=30a$ from the sector. Several observations can be made. First, for a walk anchored in the vicinity of an infinite line ($\alpha=\pi$), $\langle M^{-1}\rangle_{\vec{x}}$ appears to be constant, i.e., independent of $x$ and equal to the factor found in the homogeneous case (section \ref{sec:homo}). Thus we see that when the walk starts at a distance much greater than the lattice constant, the modification we propose to the entropic pressure is independent of the starting position of the walk. In this case the correction is reduced to a multiplicative factor which depends on a small number of parameters such as the type of lattice, the dimension of the system and the universality class of the walks (e.g. RWs versus SAWs). However, from Fig. \ref{fig:correctionFactor} we also see that $\langle M^{-1}\rangle_{\vec{x}}$ does depend on the shape of the surface, and varies when we approach the tip of the sector. For $\alpha=3\pi/2$, when the polymer is held outside of a $\pi/2$ sector (i.e., inside of a $3\pi/2$ sector), $\langle M^{-1}\rangle_{\vec{x}}$ increases near the corner, whereas for $\alpha=\pi/2$, when the polymer is confined inside the $\pi/2$ sector, $\langle M^{-1}\rangle_{\vec{x}}$ decays to zero at the corner. Note that the pressure behaves in a similar way (Fig. \ref{fig:RWsectorPressure}). The behaviour of $\langle M^{-1}\rangle_{\vec{x}}$ near the corners can be understood qualitatively in the following way: When the polymer makes contact with the surface in an area that is increasingly confined (e.g. close to the corner when $\alpha=\pi/2$), we can expect that more contacts were made on the surface in nearby points since the polymer has fewer options to escape into the bulk. Thus, the total number of contacts with the surface, $M$, will be larger for these configurations, and the  factor $\langle M^{-1}\rangle_{\vec{x}}$ will be smaller. Note that in this case it is harder for the polymer to reach the confined area and therefore the entropic pressure will also be reduced in the vicinity of the corner. On the other hand, when the area in question is less confined, as in the vicinity of the corner with $\alpha=3\pi/2$, the polymer can reach it more easily, and can more easily escape into the bulk after making contact with the surface, thus making a smaller number of contacts with the boundary. Therefore, for $\alpha=3\pi/2$, the pressure and $\langle M^{-1}\rangle_{\vec{x}}$ are both increased in the vicinity of the corner.

\section{Summary and Conclusions}

Polymers on lattices are often used to study properties of polymers in continuous space, even though there are important differences in the definition and behavior of physical properties in lattice and continuous systems. When dealing with macroscopic quantities such as the force acting on a large object, the results in the lattice systems match those in continuous space. However, this is not always the case when dealing with local properties such as the entropic pressure on a boundary surface. The natural way to define the pressure $P(\vec{x})$ at the point $\vec{x}$ along the boundary in a lattice system is as the change in the free energy of the system which results from excluding the point $\vec{x}$, divided by the volume element related to $\vec{x}$. However, this definition of the pressure is inadequate when we want to use the lattice polymer models to represent polymers in continuous space, where an important requirement is that when the pressure is integrated over the entire surface, the result should be the total force $F$ acting on the surface. It is known\cite{Jensen2013} that $P(\vec{x})$ defined above does not satisfy this condition.

For a polymer in a large box, where the pressure on the wall is constant, we show that the difference between the integral of the pressure and the total force is a constant factor that can be calculated from the statistics of polymers in free space. Thus in this case this surface effect is in fact a bulk property,\cite{Wang2008,Wang2008a} that depends on a small number of parameters such as the type of lattice and the universality class of the lattice polymer (e.g. RWs versus SAWs).

For non-homogeneous cases, we define a modified entropic pressure, denoted $\bar{P}$ in dimensionless units, that, upon integration, {\em does} result in the correct total force acting on the surface. Even though in many cases, when the surface geometry is simple, it is sufficient to divide the lattice pressure by a constant value as mentioned above, our modified pressure is guaranteed to result in the correct total force after integration even on more complicated boundaries. Note that the total force matches the one measured in continuous space. Also, computing our modified pressure in simulations does not require significant numerical effort.

We show that our modified pressure calculated for RWs on a square lattice near scale invariant repulsive boundaries matches the exact results obtained for continuous polymers in this geometry, and use this system to study the properties of the proposed modification. We show that it is {\em non-local}, i.e., it depends on long range correlations between contact points along the surface. We conclude that the entropic pressure of a lattice polymer cannot be considered as a purely local property that results in the total force after integration, like the pressure in the continuous systems.

\begin{acknowledgments}
We thank M. Kardar for useful discussions.
This work was supported by the Israel Science Foundation grant 186/13.
\end{acknowledgments}

\bibliographystyle{apsrev}
\bibliography{LatticePressureBib}
\end{document}